\documentclass[runningheads]{llncs}
\usepackage{geometry}
\usepackage [autostyle, english = american]{csquotes}
\MakeOuterQuote{"}
\newenvironment{myfont}{\fontfamily{pcr}\selectfont}{\par}
\usepackage{graphicx}
\usepackage{hyperref}
\usepackage{float}
\usepackage{listings}
\usepackage{color}
\begin{document}
\title{QualiBD Tool: Implementation Details}
%
%
\author{Darlan Arruda\orcidID{0000-0002-6756-2281}}
\authorrunning{Arruda, D.}
%
\institute{University of Western Ontario \\
Department of Computer Science, London, Canada \\
\email{darruda3@uwo.ca}}

\maketitle              
\begin{abstract}
This paper describes the tools, technologies, and frameworks used in the implementation of the QualiBD, a tool for modelling quality requirements for Big Data Software Applications. 

\end{abstract}
\section{Introduction}
\label{section:intro}
The Big Data paradigm has introduced a new set of attributes (e.g., volume, velocity, variety, and veracity), that along with the traditional systems quality attributes (e.g., performance, reliability, security, etc.) pose significant challenges to the design and development of Big Data software applications  \cite{Nazim}\cite{Noorwali}. Therefore, in the field of RE, one of the research challenges is to be able to integrate these complementary set of attributes in the analysis, specification, and modelling of system requirements \cite{Nazim}\cite{arruda18}. However, the Big Data characteristics are rarely discussed in conjunction with the traditional systems' quality attributes in Big Data systems building \cite{Noorwali}.

As part of ongoing research - built upon the concepts of the NFR Framework by Chung et al., \cite{Chung} - a goal-oriented requirement modelling language is proposed. The language enables users to model quality requirements for Big Data applications that incorporate both Big Data characteristics and systems' quality attributes on the same requirement representation through what is called \textit{Permutation} (e.g., Velocity x Latency) \cite{Noorwali}. In the proposed language, requirements are expressed as \textit{Goals}. Quality attributes are expressed as\textit{ NFR Soft-goals}. \textit{Goals} have \textit{Big Data characteristics} and \textit{NFR Soft-goals} associated to them from which  \textit{Permutations} are defined. \textit{Permutations} have \textit{Permutation Attributes} (e.g., quantitative and qualitative information) to further characterise them. Based on the identified Goals and their associated \textit{Permutations}, solutions (expressed as \textit{Operationalising Soft-goals}) are proposed to fulfil a given requirement. Each proposed solution should be supported by a rationale (expressed in the proposed language as \textit{Claim Soft-goals}).

In \cite{arruda19}, we described the QualiBD, a modelling tool that realises the aforementioned goal-oriented Big Data requirements language allowing for the modelling of quality requirements for Big Data software applications. In this paper, then, we describe the technologies  and  frameworks  used  in  the  implementation  of the QualiBD Tool.

This paper is organised as follows: next section overviews the tool implementation process. Section 3 presents the QualiBD Tool graphical user interface and warning messages. Finally, Section 4 summaries this paper.

\section{Tool Implementation}
\label{section:toolImplementation}
The implementation of the QualiBD tool consisted of two major steps: (i) modelling and code generation; and (ii) graphical editor definition. 
For the modelling and code generation, we used the Eclipse Modelling Framework (EMF), a modelling framework and code generation facility for building tools and applications based on a structured data model \cite{EMFBook}. For the graphical portion of the tool, we used Sirius \cite{SiriusDoc}, an Eclipse project that allows for the creation of graphical modelling tools by leveraging the Eclipse modelling technologies such as EMF and the Graphical ModellingFramework (GMF).

On the EMF side, we define the domain model and create a concrete instance of that model that is dynamically interpreted using a runtime within the Eclipse IDE environment. On the Sirius side (on the\textit{ Sirius Specification Editor}), we define the modelling tool - composed of all modelling elements, behaviour, java services, expressions, and navigation tools. The modelling tool references domain model defined in EMF. The Graphical representation of the model is created using the defined modelling tool. The graphical representation represents the concrete instance of the defined domain model. The concrete instance of the domain model conforms with the domain model defined in EMF.

\subsection{Modelling and Code Generation}
\label{section:modelling}
\begin{figure}
\includegraphics[width=\textwidth, height=7.2cm]{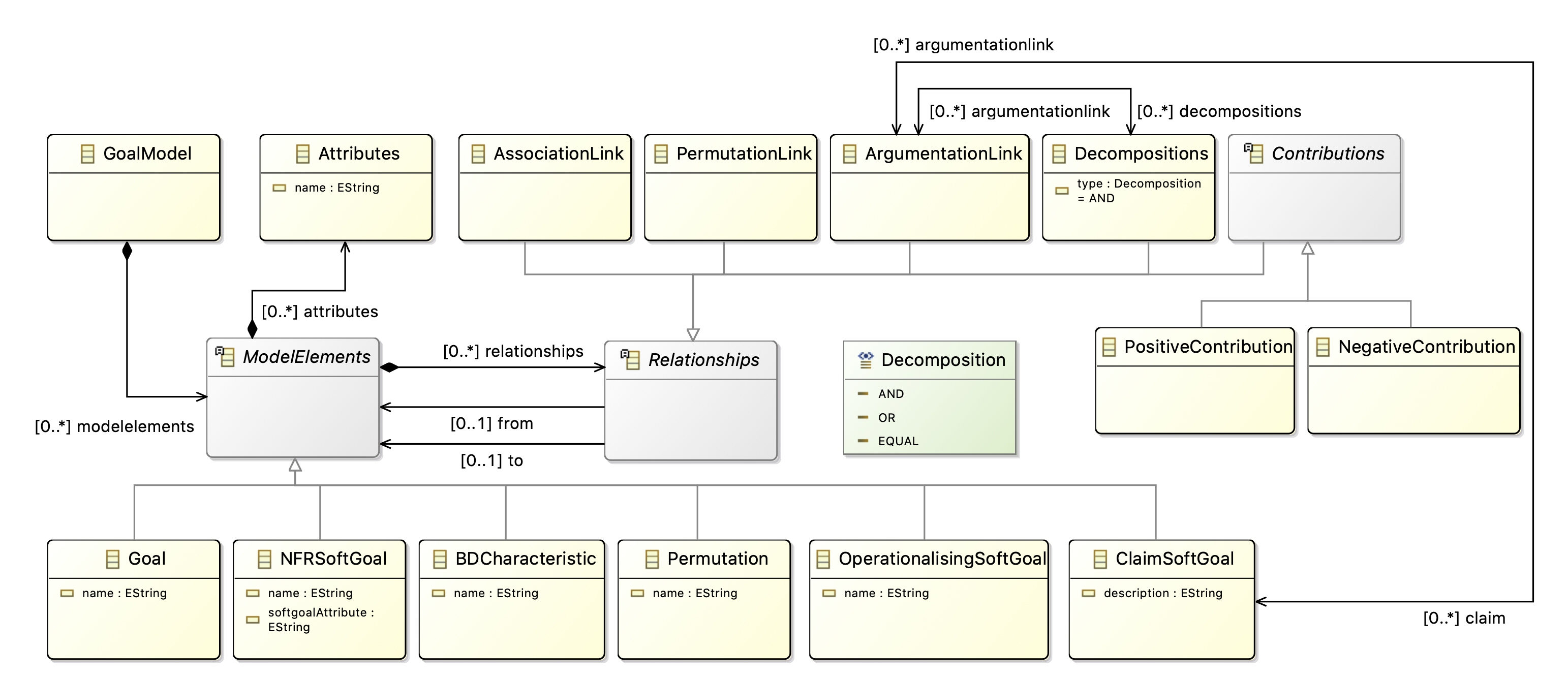}
\caption{Ecore Meta-model of the QualiBD Tool.} 
\label{metamodel}

\end{figure}

The implementation of the QualiBD tool started with the definition of a domain model that describes the modelling elements using the EMF. The model used to represent models in EMF is called Ecore \cite{EMFBook}. An Ecore can be considered a subset of a UML class diagram \cite{EMFDoc}. EMF allows for the modelling of meta-class (EClass), packages, and several different types of references (EReferences) such as compositions and inheritance. An EClass can contain different attributes and operations. An attribute (EAttribute) has a data type (EDataType) which can be primitive (e.g., int, float, boolean,) or object type (e.g., a class) \cite{EMFBook}. Figure \ref{metamodel} depicts the Ecore meta-model of the QualiBD Tool.

\subsection{Graphics Editor}
\label{section:graphicaleditor}
For the definition of the graphical portion of the QualiBD tool, we used Eclipse Sirius. The Sirius official documentation \cite{SiriusDoc} states that a modelling workbench created with Sirius is composed of a set of Eclipse editors (such as diagrams, tables and trees) which enables users to create, edit and visualise EMF models. The editor which summarizes the complete structure of the modelling workbench, its behaviour, and all the edition and navigation tools is dynamically interpreted by a runtime within the Eclipse environment  \cite{SiriusDoc}.

Before diving into the steps taken during the definition of the graphical portion of the QualiBD Tool,  let's  discuss  some  of  the concepts underlying Eclispe Sirius \cite{SiriusDoc}.

The main concepts in Sirius are (based on the official Sirius documentation \cite{SiriusDoc}): \textit{(i) Viewpoint}: represents a set of representation specifications and extensions. It is considered one of the core elements of Sirius; \textit{(ii) Representation}: set of graphical elements that represent the domain data, in order words, the concrete instance of the Ecore metamodel; \textit{(iii) Mappings}: identifies the sub-set of semantic model elements that would appear in the representation. It is also used to indicate how they should be represented; \textit{(iv) Styles}: used to customize the appearance of the defined elements; \textit{(v) Tools}: used to add edition capabilities to the graphical editor allowing end-users to create, edit, and delete model elements.

Additionally,  when defining model elements, edges, and tools in Sirius, we will be using some required interpreted expressions to configure them. These can be queries to select elements or more general-purpose expression to compute a value, for instance \cite{SiriusDoc}. The recommended language for writing queries and expressions in Sirius is the Acceleo Query Language (AQL). It is also used to navigate and query an Ecore model defined in EMF \cite{AQLDoc}. However, Sirius also supports other common expression interpreters \cite{SiriusDoc} such as (i) Var: provides direct access to the value of a named variable; (ii) Feature: offers direct access to a named feature of the current element. For example, instead of \textit{aql:self.name}, the equivalent using the Feature interpreter would be \textit{feature:name}; and (iii) Service: can used to invoke a service method (e.g., Java services) on the current element.

\begin{figure}
\centering
\includegraphics[width=12cm, height=2.4cm]{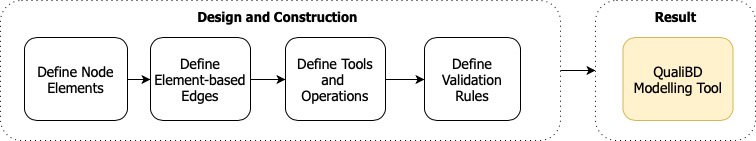}
\caption{Steps taken in the definition of the graphical editor portion of the QualiBD Tool.} 
\label{Phases}
\end{figure}

In the next subsections, we describe the steps followed in order to define the graphical editor portion of the QualiBD Tool in Sirius (as depicted in Figure \ref{Phases}). The steps represented within the "Design and Construction" phase are defined in the Sirius Specification Editor as described in the beginning of Section \ref{section:graphicaleditor}.

\subsubsection{Defining Node Elements}
When creating a node, we must describe which model element will be displayed by the modelling tool.  A model element can be displayed via either an image or a geometric shape.  For that, we must specify the following properties:

\begin{itemize}
    \item \textbf{ID}:  this is the unique identifier of the element we are defining.
    \item \textbf{Domain class}:  defines the type of element represented by the node we are creating. For instance, consider: \textit{bigDataModelling::Goal} where \textit{bigDataModelling} is the namespace (NS)  prefix,  in  order  words,  the  name  of the  Ecore  meta-model  and \textit{Goal} is  the  name  of  the element  (Eclass)  in  that  meta-model.  Specifying the NS is important to prevent eventual conflicts with another metamodel that could define a class of the same name.
    \item \textbf{Semantic candidate expression}: Restricts to the list of elements to consider before creating the graphical elements. If not set, then all semantic models in session will be browsed and any element of the given type validating the precondition expression will cause the creation of the element. If we set this attribute then, only the elements returned by the expression will be considered. For instance, in the case of our modelling tool, we feature the \textit{modelelements} abstract class where it is the class that extends the goal class and other elements specified in this modelling language.
\end{itemize}

\subsubsection{Defining Element-based Edges}
Edges in Sirius can be defined as relation-based or element-based \cite{SiriusDoc}. Relation-based edges are used to represent a relation between model elements such as containment or references whereas element-based edges are used when a semantic model element exists to represent the relation itself \cite{SiriusDoc}. Since all the relationships in the QualiBD tool are represented semantically through classes in the Ecore model, we only used element-based edges. Figure \ref{Edgecreation} depicts the properties defined to create the Permutation Link element-based edge. With reference to Figure \ref{Edgecreation}:

\begin{figure}[ht]
\centering
\includegraphics[width=8cm, height=3.8cm]{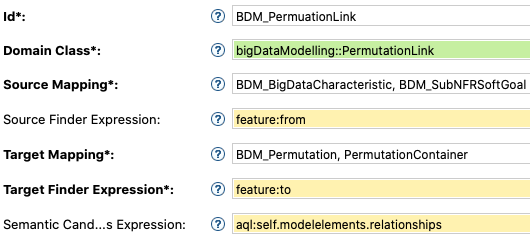}
\caption{Properties of an element-based relation for the \textit{Permutation Link} element in Sirius.} \label{Edgecreation}
\end{figure}

\newpage
 \begin{itemize}
     \item 	\textbf{ID}: this is the unique identifier of the element we are defining. 
     \item  \textbf{Domain Class}: the name of the domain class that triggers the creation of the new edge. In the context of Figure 2, the Permutation Link class within the bigDataModelling metamodel. Again (repeated for convenience), the specification of the NS is important to prevent eventual conflicts with another metamodel that could define a class of the same name.
     \item \textbf{Source Mapping}: maps the element from where the edge should start.
     \item \textbf{Source Finder Expression and Target Finder Expression}: will be evaluated in the context of the semantic element of the edge. It should return the actual elements that the edge connects.
     \item \textbf{Target mapping}: maps the element from which the edge should end.
     \item \textbf{Semantic candidate expression}: Restricts to the list of elements to consider before creating the graphical elements. We use an AQL expression that points the to the \textit{Relationships} abstract class defined in the QualiBD Ecore meta-model. Only the elements returned by the expression will be considered. AQL expression used: \begin{myfont}\textit{aql:self.modelelements.relationships.}\end{myfont}
     \end{itemize}
\subsubsection{Defining Tools and Operations}
Once all the elements (nodes and edges) are defined, we can establish the tools that will be displayed in the palette of Eclipse, that turn, will allow the end-user to create and edit new model elements onto the tool container (canvas). Without tools, the models would be "visualisations only", without any edition capabilities \cite{SiriusDoc}. 

\begin{figure}[ht]
\centering
\includegraphics[width=7.7cm, height=4.9cm]{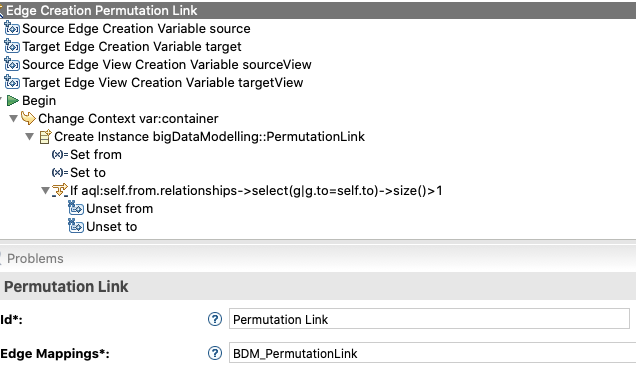}
\caption{Properties of an edge creation tool in Sirius for the \textit{Permutation Link} element.} \label{tools}
\end{figure}

In the QualiBD Tool, the following types of tools were defined \textit{(i) Element Creation}; and \textit{(ii) Element Edition}. The former, enables the creation of instances of model elements. The later, adds editing capabilities in the QualiBD Tool. Example of editing capabilities supported by the QualiBD tool are: \textit{(a) Direct Edit Label} that allows for the modification a graphical object label (e.g, name of a goal model element) direct from the graphical diagram; and \textit{(b) Reconnect Edges} that allows end-users to change the source and/or target of an edge by moving the corresponding end onto another graphical model element \cite{SiriusDoc}. As an example - Figure \ref{tools} depicts the properties - \textit{of the Element Creation Tool} - defined to create the edge tool Permutation Link. Please note that the procedure for creating node tools is similar to edge tools, thus, only one example will be provided.

\newpage
With reference to Figure \ref{tools}:

\begin{itemize}
    \item The \textit{Begin} element has no property. It serves as an entry point to the specification of the behavior of our tool.
    \item Within the \textit{Begin} element, we define the \textit{Change Context} operation, which serves to changes the context to a new element and executes any contained sub-operations\cite{SiriusOperations}. We also define the \textit{Create Instance} operation that is used to create new semantic elements to be added into the end-user’s model. For that, we must specify the \textit{Type Name} (using the same syntax as for \textit{Domain Class} properties) of the new object to be created and the Reference Name through which the created element will be attached to \cite{SiriusOperations}.
    \item For edge tools (in the context of QualiBD tool), we also specify the Set operation that, as the name says, is used to set the value of a feature. It can be an attribute or a reference of the current element. In the case of the \textit{Permutation Link} edge tool, the source and the destination of the relation defined in our Ecore metamodel (in our case the relations \textit{From} and \textit{To}).
    \item Additionally, we can define some operations to control the behaviour of the tool being created. In the context of the
    all relation-based elements (such as association, permutation, decomposition, argumentation, and contribution links) defined in the QualiBD tool, we set an operation to prevent the user to establish duplicated connections amongst model elements. For that, we used the \textit{If} operation that evaluates its \textit{Condition Expression}.  An example of a condition expression in AQL used by the \textit{If} operation is:
    
    \begin{myfont}\textit{aql:self.from.relationships->select(g|g.to=self.to)->size()>1.} \end{myfont}If the expression (interpreted as a boolean)  returns "false", If does nothing. If it returns "true", then it executes any sub-operations defined under the \textit{If} operation in the order of definition \cite{SiriusOperations}.  In the case of the QualiBD tool, it "cancels" (unsets) the creation of edges between two nodes when the relation already exists in the representation. 
\end{itemize}

\subsubsection{Define Validation Rules}
In order to allow the QualiBD tool to function properly and minimise possible user induced errors, we specified some elementary validation features focused on the completeness and accuracy of the models created The validation features are briefly described as follows: 

\begin{itemize}
    \item \textbf{Empty Labels}: This rule checks for the existence of model elements with empty labels. For that, we specified a semantic validation rule. Eclipse Sirius offers three levels of semantic validation rules: Message, Warning, and Error. The semantic validation rule is characterised by an \textit{audit expression}. If the audit expression returns \textit{True}, then nothing happens. If the audit expression returns \textit{False}, then a validation issue will be pointed out (e.g., a warning message to be displayed to the end-user in the \textit{Problems} view of Eclipse). An example of an audit expression used is: \begin{myfont}\textit{aql:self.name<>null.}\end{myfont}
    
    \item \textbf{Model Elements with Empty Connections}: This rule checks for the existence of model elements with empty connections (\textit{refinements}). For that, we define the same procedures described in the "Empty Labels Validation". However, for this one specifically, instead of using an AQL expression in the audit expression definition, we invoked a Java service method that navigates the existing model elements on the tool canvas, and checks whether there are established connections amongst them. If empty connections are found, it returns the validation warning message. Otherwise, it returns \begin{myfont}\textit{null}.\end{myfont}

\end{itemize}

\section{QualiBD Tool Graphical User-Interface}

As previously  described, the QualiBD Tool (a Sirius modeller) runs as a plug-in on the Eclipse environment, thus , its graphical user-interface is the default one provided by the Eclipse IDE. Figure 5 provides an overview of the tool running on Eclipse.

\begin{figure}[ht]
\begin{center}
\includegraphics[height = 7.9cm, width=\textwidth]{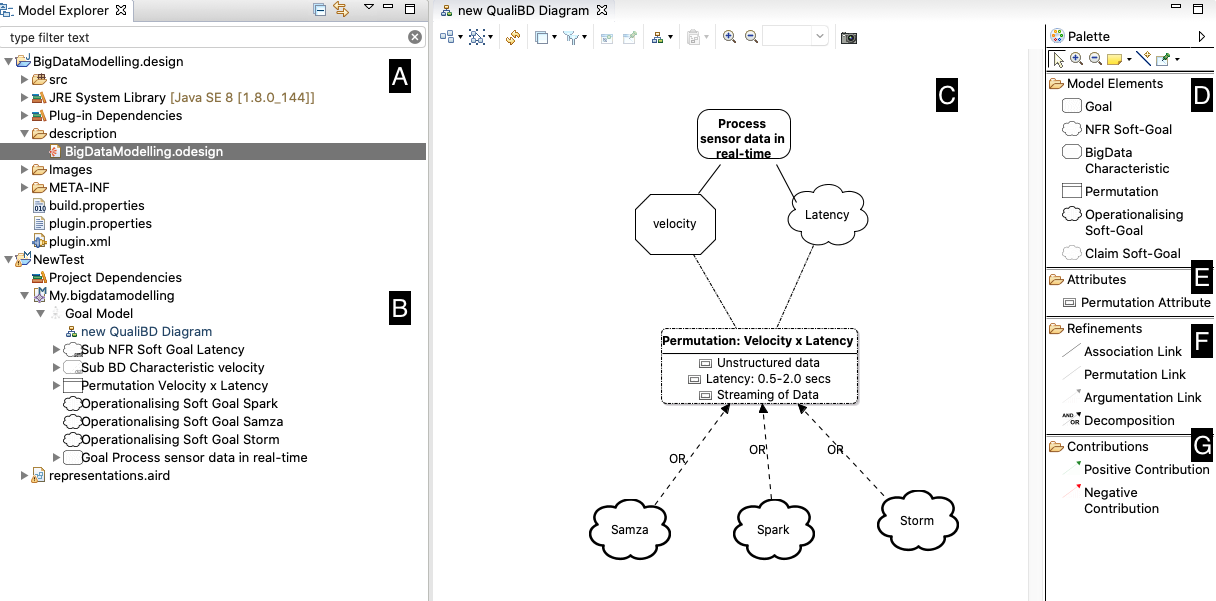}
\caption{QualiBD Tool Graphical User-Interface \cite{arruda19}.}
\end{center}
\label{interfaceTool}
\begin{minipage}{1\textwidth} 
{\footnotesize \textbf{This figure depicts}: A) Sirius Design project where the nodes, tools, and behaviour attributes of the modelling tool are defined; B) Eclipse project that creates a concrete instance of the defined Ecore domain-model; C) Tool canvas where models can be created, edited and deleted; D) Portion of the palette tool that allows end-users to create instances of model elements; E) Portion of the palette tool that allows end-users to add permutation attributes to permutation containers; F) and G) Portions of the palette tool that allow end-users to define the types of refinements (relations) supported by the QualiBD tool.\par}
\end{minipage}

\end{figure}

\newpage
\section{Summary}
In  this  paper,  we  described the implementation of  the  QualiBD,  a  tool for  modelling  quality  requirements  for  Big  Data  Software applications.  The  definition  of  our tool  is  comprised  of  two steps:  (i)  domain  model  definition  and  code  generation,  and(ii)  graphical  editor  definition.  For  that,  we  used  eclipse modelling and Sirius Frameworks, respectively.  On the EMF side, we defined the tool's structured data model. On the Sirius side, we defined the graphical editor behaviour. The graphical editor portion of the tool was defined in four incremental steps: (i) definition of node elements, (ii) definition of edge elements, (iii) definition of tools and operations, and (iv) definition of validations rules. Although the tool is still in its early stages, it successfully enables end-users to graphically model quality requirements for big data applications, thus, aiding in more complete Big Data requirements specifications.

\subsubsection{Acknowledgements.}This research is supported, in part, by grants from CNPq, The National Council of Technological and Scientific Development – Brazil and NSERC, Natural Science and Engineering Research Council of Canada.

%
%
%

\end{document}